\documentclass[sigconf]{acmart}
\usepackage{pifont}
\usepackage{graphicx}  
\usepackage{float}  
\usepackage{subfigure} 
\usepackage{algorithm}
\usepackage{algorithmicx}
\usepackage{algpseudocode}
\usepackage{amsmath}
\usepackage{enumerate}
\usepackage{enumitem}
\usepackage{multirow}

\AtBeginDocument{%
  }

\copyrightyear{2024}
\acmYear{2024}
\setcopyright{rightsretained}
\acmConference[KDD '24]{Proceedings of the 30th ACM SIGKDD Conference on Knowledge Discovery and Data Mining}{August 25--29, 2024}{Barcelona, Spain}
\acmBooktitle{Proceedings of the 30th ACM SIGKDD Conference on Knowledge Discovery and Data Mining (KDD '24), August 25--29, 2024, Barcelona, Spain}\acmDOI{10.1145/3637528.3671519}
\acmISBN{979-8-4007-0490-1/24/08}

\makeatletter
\gdef\@copyrightpermission{
 \begin{minipage}{0.3\columnwidth}
  \href{https://creativecommons.org/licenses/by/4.0/}{\includegraphics[width=0.90\textwidth]{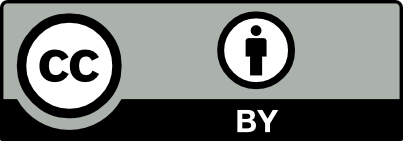}}
 \end{minipage}\hfill
 \begin{minipage}{0.7\columnwidth}
  \href{https://creativecommons.org/licenses/by/4.0/}{This work is licensed under a Creative Commons Attribution International 4.0 License.}
 \end{minipage}
 \vspace{5pt}
}
\makeatother

\begin{document}

\title{Unified Dual-Intent Translation for Joint Modeling of Search and Recommendation}
\author{Yuting Zhang}\authornote{Yuting Zhang and Yiqing Wu are also at University of Chinese Academy of Sciences}
\affiliation{%
  \institution{Institute of Computing Technology, \\
  Chinese Academy of Sciences}
  \city{Beijing}
  \country{China}}
\email{zhangyuting21s@ict.ac.cn}

\author{Yiqing Wu}\authornotemark[1]

\affiliation{%
  \institution{Institute of Computing Technology, \\
  Chinese Academy of Sciences}
  \city{Beijing}
  \country{China}}
\email{iwu_yiqing@163.com}

\author{Ruidong Han}
\affiliation{
  \institution{Meituan}
  \city{Beijing}
  \country{China}}
\email{hanruidong@meituan.com}

\author{Ying Sun}
\affiliation{
  \institution{Thrust of Artificial Intelligence, The Hong Kong University of Science and Technology (Guangzhou)}
  \city{Guangzhou}
  \country{China}}
\email{yings@hkust-gz.edu.cn}

\author{Yongchun Zhu}
\affiliation{
  \institution{Institute of Computing Technology, \\
  Chinese Academy of Sciences}
  \city{Beijing}
  \country{China}}
\email{zhuyc0204@gmail.com}
\author{Xiang Li}
\author{Wei Lin}
\affiliation{
  \institution{Meituan}
  \city{Beijing}
  \country{China}}
\email{lixiang245@meituan.com}
\email{linwei31@meituan.com}
\author{Fuzhen Zhuang}
\authornote{Fuzhen Zhuang and Zhulin An are Corresponding authors.}
\affiliation{%
  \institution{Institute of Artificial Intelligence, Beihang University}
  \city{Beijing}
  \country{China}}
  \affiliation{
      \institution{
        Zhongguancun Laboratory
      }
      \city{Beijing}
      \country{China}
  }
\email{zhuangfuzhen@buaa.edu.cn}

\author{Zhulin An}
\authornotemark[2]
\author{Yongjun Xu}
\affiliation{%
  \institution{Institute of Computing Technology, Chinese Academy of Sciences}
  \city{Beijing}
  \country{China}
}
\email{anzhulin@ict.ac.cn}
\email{xyj@ict.ac.cn}
\renewcommand{\shortauthors}{Yuting Zhang et al.}

\begin{abstract}
Recommendation systems, which assist users in discovering their preferred items among numerous options, have served billions of users across various online platforms. Intuitively, users’ interactions with items are highly driven by their unchanging inherent intents (e.g., always preferring high-quality items) and changing demand intents (e.g., wanting a T-shirt in summer but a down jacket in winter). However, \textbf{both types of intents are implicitly expressed in recommendation} scenario, posing challenges in leveraging them for accurate intent-aware recommendations. Fortunately, \textbf{in search} scenario, often found alongside recommendation on the same online platform, \textbf{users express their demand intents explicitly through their query words}. Intuitively, in both scenarios, a user shares the same inherent intent and his/her interactions may be influenced by the same demand intent. It is therefore feasible to utilize the interaction data from both scenarios to reinforce the dual intents for joint intent-aware modeling. But the joint modeling should deal with two problems: (1) \textbf{accurately modeling users' implicit demand intents in recommendation}; (2) \textbf{modeling the relation between the dual intents and the interactive items}. To address these problems, we propose a novel model named \underline{U}nified \underline{D}ual-\underline{I}ntents \underline{T}ranslation for joint modeling of \underline{S}earch and \underline{R}ecommendation (UDITSR). To accurately simulate users' demand intents in recommendation, we utilize real queries from search data as supervision information to guide its generation. To explicitly model the relation among the triplet <inherent intent, demand intent, interactive item>, we propose a dual-intent translation propagation mechanism to learn the triplet in the same semantic space via embedding translations. Extensive experiments demonstrate that UDITSR outperforms SOTA baselines both in search and recommendation tasks. Moreover, our model has been deployed online on Meituan Waimai platform, leading to an average improvement in GMV (Gross Merchandise Value) of 1.46\% and CTR(Click-Through Rate) of 0.77\% over one month. 
\end{abstract}



\maketitle
\section{Introduction}
Aiming to help users discover items of interest from a vast array of options,
recommendation systems have become an essential component of various online platforms, such as e-commerce~\cite{zhou2018deep,zhou2019deep,pi2020search} and digital news services~\cite{li2010contextual,covington2016deep,tai2021user}. Existing recommendation models~\cite{hu2008collaborative,zhou2018deep,he2017neural,zhou2019deep} typically exploit users' implicit feedback, such as click history, to predict their interests. For instance, traditional Collaborative Filtering (CF)~\cite{hu2008collaborative} assumes that users will interact with items similar to those with which they've previously interacted. Furthermore, various models~\cite{zhou2018deep,zhou2019deep} have been developed to capture the sequential dynamics of users' implicit feedback to model their evolving interests.

\begin{figure}
    \centering
    \includegraphics[width=0.6\linewidth]{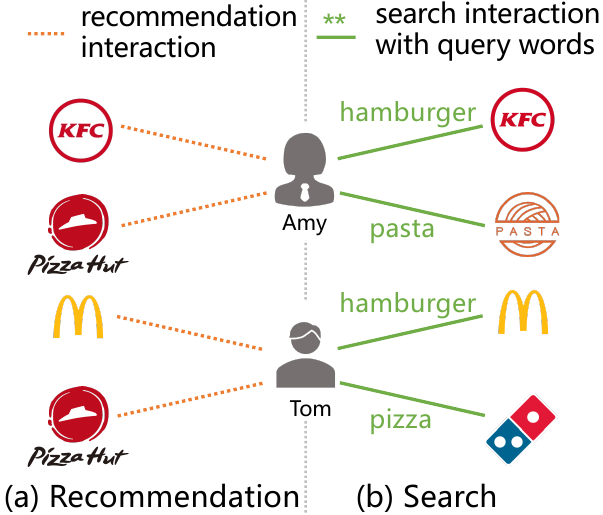}
    \caption{Examples of interaction behaviors in recommendation and search scenarios.}
    \label{fig:dual_scene}
\end{figure}

In practice, user feedback patterns in recommendation systems are highly driven by their complex intents, which can be broadly categorized into unchanging inherent intents and changing demand intents. For example, Amy and Tom may have the same noodle demand but choose different restaurants due to Amy's inherent intent for spicy flavors and Tom's for sweet. Besides, a single user's interactions can vary due to their changing demands. Yet, these intents are often implicitly expressed in the recommendation, presenting a challenge for accurate intent-aware recommendations. 
Existing intent-aware recommendation models~\cite{chen2019air,zhu2020sequential,liu2020basket} typically rely on users' implicit feedback to learn their intents. However, these models encounter a significant problem: different users may have different inherent or demand intents despite similar historical feedback.  As shown in Figure~\ref{fig:dual_scene}(a), Amy's interaction with Pizza Hut might indicate a demand intent for pasta, while Tom may demand pizza instead. 
Ideally, recommendation systems should suggest pasta-related options to Amy and pizza-related ones to Tom. However, without any explicit intent information, existing models struggle to distinguish between these intents,  resulting in inaccurate recommendations. 

Fortunately, in search services, which often accompany recommendation services on the same online platform, users explicitly express their demand intents through query words, as shown in Figure~\ref{fig:dual_scene}(b). Such explicit search demand information can serve as additional explicit information to assist in learning implicit demand intents for recommendation.  Indeed, both search and recommendation tasks aim to comprehend users' intents to aid them in obtaining desired items~\cite{belkin1992information}. 
In addition, in search scenario, users' interactions are influenced not only by their explicit demand intents but also by their personalized inherent intents.  Yet, search models typically focus on the match between search results and users' demand intents, often overlooking the impact of their personalized inherent intents, which are indeed significant~\cite{sondhi2018taxonomy}. Intuitively, in both scenarios, a user maintains the same inherent intent and his/her behaviors are likely to be determined by the same demand intent. Therefore, it is feasible to leverage interaction data from both scenarios to reinforce or complement each other's dual intents for joint intent-aware modeling. Nevertheless, this joint modeling is not trivial due to the following challenges:

(1) \textbf{How to accurately model a user's implicit demand intent in recommendation with search data?} 
A user's demand intent is implicit within recommendation but is explicitly indicated by search queries. If the changing demand intents in recommendation can be accurately generated, search and recommendation can be well modeled in a unified manner. The existing method, SRJGraph~\cite{zhao2022joint}, employs the unchanging padding query in recommendation for unified modeling.  This approach assumes an unchanging demand intent across all recommendation interactions, which may hinder recommendation performance. To learn demand intents, an intuitive approach is to simply incorporate users' historical queries as additional demand information into the recommendation model. However, without explicit supervision to verify the accuracy of demand intents, there may be a significant discrepancy between the learned and the actual demand intents.

(2) \textbf{How to couple the dual intents to model the relation among the intents and the interactive items?}
Both inherent intent and demand intent affect the interactive item. 
Intuitively, the superimposition of inherent intents (e.g., preferring \textit{cheap }items) and changing demand intents (needing a \textit{T-shirt} in summer but a \textit{down jacket} in winter) leads to changing interactive results (interacting with a \textit{cheap T-shirt} and \textit{cheap down jacket}, respectively). In essence, the demand intent can be regarded as the changing deviation from the inherent intent to the changing interactive item. 
A common approach is to simply feed the two intents as input features, but it cannot fully capture the relation between the dual intents and the interactive item.  

To tackle these challenges, we propose a novel model named Unified Dual-Intent Translation for joint modeling of Search and Recommendation (UDITSR). Overall, UDITSR comprises a search-supervised demand intent generator and a dual-intent translation module. Specifically, in the demand intent generator, search queries serve as supervision information, allowing us to learn and understand a user's changing demand intent for recommendations both reliably and accurately.
Moreover, we develop a dual-intent translation propagation mechanism. This mechanism explicitly models the interpretable relation among the triplet elements\texttt{--}user's <inherent intent, demand intent, interactive item>\texttt{--}within a shared semantic space by employing embedding translations. Particularly, we design an intent translation contrastive learning to further constrain the translation relation. Extensive offline and online experiments were conducted to demonstrate our model's effectiveness. To gain deeper insights into the effectiveness of our model, we also provide a visual analysis of relevant intents.

\section{Related Work}
\subsection{Recommendation and Search Models}
Recommendation aims to filter items from vast candidate pools to match user interests. Traditional models, such as Collaborative Filtering (CF), assume users with similar behaviors share item preferences~\cite{sarwar2001item,cheng2016wide,he2017neural,guo2017deepfm}. Later studies~\cite{zhou2018deep,sun2019bert4rec,feng2019deep} focus on decoding users' evolving interests from their historical behaviors, using techniques like DIN~\cite{zhou2018deep}, which employs attention mechanism to connect past behaviors with current targets. 
Recognizing that users' interactions are driven by their intrinsic intents, recent studies~\cite{wang2019modeling, chen2019air, zhu2020sequential, wang2020intention2basket} exploit users' historical behavior sequences to understand their changing intents, aiming to better meet user needs. For instance, KA-MemNN~\cite{zhu2020sequential} uses item categories from user behavior as intent proxies,  implementing memory networks for dynamic intent modeling. However, these approaches often deduce intents from interaction behaviors or directly equate behavior with intent, without mining real intrinsic intents. In contrast, our model utilizes the user's actual demand intents in the search scenario as supervision information to imitate the intents in recommendation.

Search and recommendation services often coexist on the same platform~\cite{qin2023comprehensive}. Earlier research ~\cite{belkin1992information} suggests their goals are essentially equivalent\texttt{--}helping people get the items they want, prompting studies on their joint optimization. For example, JSR~\cite{zamani2018joint} introduces a shared-parameter framework, with user and item embeddings shared. USER~\cite{yao2021user} treats recommendation behavior as a form of search behavior with unchanging padding query, unifying the modeling of search and recommendation sequences. Furthermore, SRJgraph~\cite{zhao2022joint} constructs a unified graph from search and recommendation behaviors, incorporating search queries and a padding query for recommendation as attributes of user-item edges. 
These models assume the query-related intents in recommendation are unchanging while the matching degree between the query and the candidate items significantly affects search performance.
This assumption creates a significant gap between the modeling of search and recommendation, greatly hindering the effectiveness of joint modeling approaches. Our model, however, adapts to learn personalized and changing query-related intents for distinct user-item pairs in recommendation, thus enhancing the unification of joint search and recommendation.




\subsection{Graph Neural Network}
Graph Neural Networks (GNNs)~\cite{scarselli2008graph,wu2020comprehensive} have gained tremendous attention in recent years due to their remarkable ability to process graph-structured data. For instance, Graph Convolutional Network (GCN)~\cite{kipf2016semi} employs a localized filter to aggregate information from neighbors, and Graph Attention Network (GAT)~\cite{velivckovic2018graph} leverages the attention mechanism to weigh the importance of each neighbor node during the aggregation process. Since then, numerous variants of GNNs~\cite{zhang2019heterogeneous,yan2018spatial,derr2018signed} have been proposed to tackle various types of graphs. 
Nowadays, Graph Neural Networks have shown great potential in a wide range of applications, such as recommendation~\cite{wang2019neural,he2020lightgcn,wu2022graph} and search~\cite{niu2020dual,fan2022modeling,liu2020structural} scenarios. In this work, we propose incorporating demand intents that are generated through search supervision in recommendation scenario, as well as explicitly stated search intents, into the construction of a unified graph. Specifically, these demand intents serve as the attributes of the edges connecting users and items. Moreover, the invariant node representations for a user across different interactions are used to indicate their inherent intents. Based on the graph, we propose a novel dual-intent translation propagation for unified dual intent-aware modeling.

\section{Preliminary}

Let $\mathcal{U}$ and $\mathcal{I}$ denote the universal sets of users and items in both search and recommendation scenarios. In order to distinguish these two scenarios, we define the interaction records in each scenario as follows:

\textbf{Definition 1}.\textbf{\textit{search scenario}}: In the search data $\mathcal{X}_s$, each interaction record $x_s \in \mathcal{X}_s$  can be formulated as $x_s=(u,i,q)$, which represents that user $u\in \mathcal{U}$ clicked item $i\in \mathcal{I}$  with the explicit query $q$. The query $q$ can
be segmented into several shorter terms as $q=[w_1,\cdots,w_{|q|}]$, where $w_i$ denotes the $i$-th term and $|q|$ is the number of terms in query $q$. 

\textbf{Definition 2}.\textbf{\textit{recommendation scenario}}: In the recommendation data $\mathcal{X}_r$, each interaction record $x_r \in \mathcal{X}_r$  can be formulated as $x_r=(u,i)$,  which represents user $u\in \mathcal{U}$ clicked item $i\in \mathcal{I}$ without an explicit query.

Thereby, the \textit{double-scenario graph} including all user click behaviors in both scenarios can be constructed as follows: 

\textbf{Definition 3}.\textbf{\textit{double-scenario graph}}:  Given the set of all user click behaviors in both scenarios, denoted as $\mathcal{X} = \mathcal{X}_s \cup \mathcal{X}_r$, the \textit{double-scenario graph} can be formulated as $\mathcal{G}=(\mathcal{U}\cup\mathcal{I},\mathcal{{E}}_{s}\cup \mathcal{{E}}_{r})$. Each search edge $\epsilon \in \mathcal{E}_{s}$ corresponds to a record ($u,i,q$) in $\mathcal{X}_s$, while 
each recommendation edge $\epsilon \in \mathcal{E}_{r}$ corresponds to a record ($u,i$) in $\mathcal{X}_r$.

In Figure~\ref{fig:framework}(a), there is an example of our \textit{double-scenario graph}. For instance, user $u_1$ searches for query $q_{12}$ and then clicks item $i_2$ in search scenario. Thus, an edge exists between nodes $u_1$ and $i_2$, with query $q_{12}$ assigned as an attribute of this edge. Likewise, in recommendation scenario, when user $u_1$ clicks item $i_1$, an edge also exists between user $u_1$ and item $i_1$, but without any query attribute. Based on the above definitions, the joint modeling of search and recommendation can be defined as follows:

\begin{figure*}
    \centering
    \includegraphics[width=0.95\linewidth]{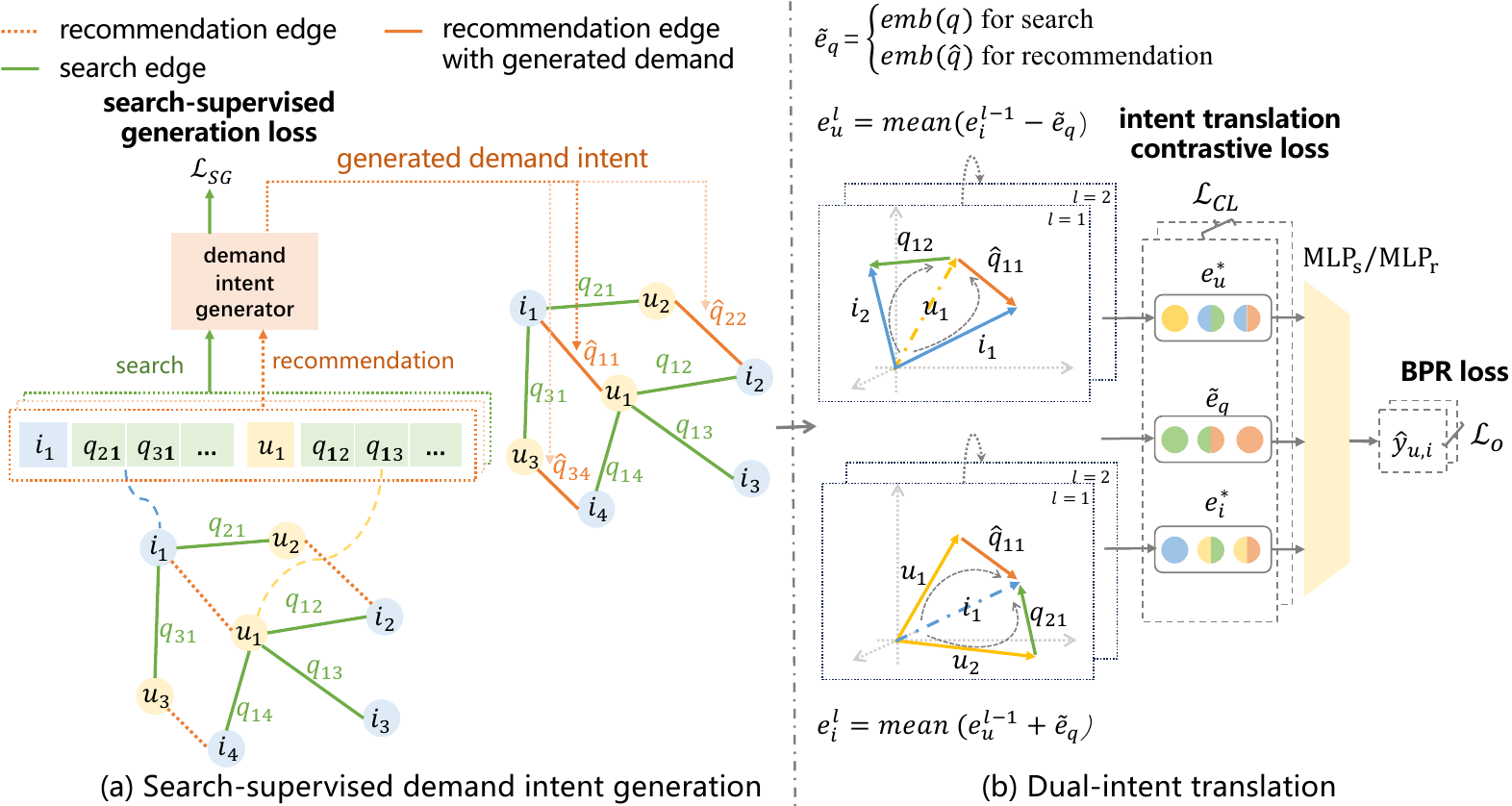}
    \caption{Overall framework of our proposed UDITSR. The \textit{mean} in \textit{dual-intent translation} represents the mean-pooling operation in Eq.~\ref{eq:mean_pooling}. For clarity, only two interaction examples are displayed for each graph aggregation in the \textit{dual-intent translation}. }
    \label{fig:framework}
\end{figure*}

\textbf{Problem definition}.
Given search data $\mathcal{X}_s$, recommendation data $\mathcal{X}_r$ and double-scenario graph $\mathcal{G}$,  this task is to train a joint model of search and recommendation to predict the most appropriate items $i\in \mathcal{I}$ that user $u\in \mathcal{U}$ will interact. 

\section{Methodology} 
In this section, we introduce UDITSR for dual intent-aware joint modeling of search and recommendation, as depicted in Figure~\ref{fig:framework}. We begin with the model's \textit{embedding layer} in Section~\ref{sec:embedding_layer}. Then, in Section~\ref{sec:generator}, we detail a \textit{search-supervised demand intent generator} that leverages search query data to infer recommendation intents, which allows us to convert the \textit{double-scenario graph} into a \textit{unified graph}. Utilizing this graph, we describe \textit{dual-intent translation propagation} to couple inherent intents and demand intents, enhanced by a contrastive loss to constrain the translation relation. Finally, the prediction layer and optimization are illustrated in Section~\ref{sec:prediction}.

\subsection{Embedding Layer}\label{sec:embedding_layer}
By feeding the user ID and item ID into the user and item embedding matrices respectively, we can obtain the embeddings of user $u$ and item $i$ as $\mathbf{e}_u,\mathbf{e}_i$. Since each query $q$ is a sequence of shorter terms as [$w_1$, $w_2$, $\cdots$, $w_{|q|}$], we can obtain the representation of query $q$ by combining the embeddings of its terms:
\begin{equation}\label{eq:query_information}
\begin{aligned}
    \mathbf{e}_q = f(\mathbf{e}_{w_1},\mathbf{e}_{w_2},\cdots,\mathbf{e}_{w_{|q|}}),
\end{aligned} 
\end{equation}
where $\mathbf{e}_{w_{k}}$ represents the embedding of the $k$-th query term in $q$ and $f(\cdot)$ denotes a combination function. In this study, we choose the element-wise sum-pooling operation because it is both efficient and effective for this combination through empirical analysis.
\subsection{Demand Intent Generation}\label{sec:generator}
\subsubsection{Search-Supervised Demand Intent Generator}
The notable difference between search and recommendation is that a user explicitly expresses demand intents in search, whereas recommendation lacks such explicit intents. To bridge this gap, we propose to utilize the abundant query information from search to supervise the generation of users' demand intents in recommendation. Below we describe the generator in detail.

Since the user's historical queries $q_u=[w^u_1,w^u_2,\cdots,w^u_{|q_u|}]$ and the item's historical queries $q_i=[w^i_1,w^i_2,\cdots,w^i_{|q_i|}]$ contain abundant demand intent information, we leverage them as auxiliary information to simulate the user's demand intent for recommendation. Similar to the processing of $q$ in Eq.~\ref{eq:query_information}, we adopt the element-wise sum-pooling operation to obtain the representation of $q_u$ as  $\mathbf{e}_{q_u} = \sum\limits_{k=1}^{|q_u|}\mathbf{e}_{w_k^u}$, where $\mathbf{e}_{w_{k}^u}$ is the embedding of the $k$-th query term in $q_u$.  Since $q_i$ contains query words from multiple users, we introduce a user-aware gate mechanism to model personalized demand intents.
Particularly, the user-aware gating network $g$ yields a distribution over the $|q_i|$ query words.  The personalized representation of  $q_i$ is then formulated as the weighted sum of the embeddings of its query words, as follows:

\begin{equation}\label{eq:itemquerygate}
    \begin{aligned}
    \mathbf{K}_{\rm g} &= \mathbf{W}_{\rm g}(\mathbf{e}_u\Vert \mathbf{e}_{w_1^u}\Vert \cdots \Vert \mathbf{e}_{w_{|q_u|}^u}),\\
    g(w_k^i) &= \frac{{\rm exp}(\mathbf{K}_{\rm g}\times {\mathbf{e}_{w_k^i}}^\top)}{\sum\limits_{k=1}^{|q_i|} {\rm exp}(\mathbf{K}_{\rm g} \times {\mathbf{e}_{w_k^i}}^\top)},\\
    \mathbf{e}_{q_i}&= \sum\limits_{k=1}^{|q_i|} g(w_k^i) \mathbf{e}_{w_k^i},
    \end{aligned}
\end{equation}
where $\cdot\Vert\cdot$ denotes the concatenation operation; $\mathbf{W}_{\rm g}$ is used to match the dimensions of vector $\mathbf{e}_{w_k^i}$ and the concatenated vector. Then, with user-related representations $\mathbf{e}_u,\mathbf{e}_{q_u}$ and item-related representations $\mathbf{e}_i,\mathbf{e}_{q_i}$, the user's demand intent about the item can be estimated as follows:
\begin{equation}
\begin{aligned}
    \hat{\mathbf{e}}_q={\rm MLP} (\mathbf{e}_u\Vert \mathbf{e}_i\Vert \mathbf{e}_{q_u}\Vert \mathbf{e}_{q_i}),
\end{aligned} 
\end{equation}
where $\rm MLP$ denotes a multi-layer perceptron. Since the ground truth queries in search data serve as the supervision information for generating demand intent, we design the generation loss as follows:
\begin{equation}
    \begin{aligned}
        \mathcal{L}_{SG}= \sum\limits_{(u,i,q) \in \mathcal{X}_s}(\mathbf{e}_q-\hat{\mathbf{e}}_q)^2.
    \end{aligned}
\end{equation}

\subsubsection{Unified Graph}
After generating the demand intents, each recommendation record ($u,i$) in $\mathcal{X}_r$ can be converted into a triplet ($u,i,\hat{q}$), where the embedding of $\hat{q}$ corresponds to the generated intents $\hat{\mathbf{e}}_q$. For simplicity, \textbf{we directly generate the representation of intent $\hat{\mathbf{e}}_q$ instead of indirectly predicting the specific query $\hat{q}$}. With the generated demand intents, the \textit{double-scenario graph} can be converted into a \textit{unified graph}.
Specifically, an additional attribute $\hat{q}$ is attached to each recommendation edge ($u,i$) in $\mathcal{G}$. For brevity, we use $\widetilde{q}/\widetilde{\mathbf{e}}_q$ to uniformly represent the real $q/\mathbf{e}_q$ in search scenario and the generated $\hat{q}/\hat{\mathbf{e}}_q$ in recommendation scenario correspondingly. Based on the unified graph, we implement the unified modeling of recommendation and search below.
\subsection{Dual-Intent Translation Propagation}
To explicitly model the relation among the dual intents and the interactive items, we propose a dual-intent translation module inspired by the triplet-based representation learning in knowledge graphs~\cite{bordes2013translating}. Specifically, we use the user's embedding representation, which remains inherent for a single user, to represent their inherent intent. The search query representation and the generated demand intent in recommendation represent the user's demand intent. The representation of an interactive item is given by its embedding. We assume that a user's changing interactive item should be close to their inherent intent plus changing demand intent.  
Consequently, we aggregate the neighbor embeddings as follows:
\begin{equation}\label{eq:mean_pooling}
\begin{aligned}
    \mathbf{e}_{i}^l &= mean\_pooling(\{\mathbf{e}_{u}^{l-1}+\widetilde{\mathbf{e}}_q, \forall u \in \mathcal{N}_i\}), \\
    \mathbf{e}_{u}^l &= mean\_pooling(\{\mathbf{e}_{i}^{l-1}-\widetilde{\mathbf{e}}_q, \forall i \in \mathcal{N}_u\}),
\end{aligned} 
\end{equation}
where $\mathcal{N}_u$ and $\mathcal{N}_i$ denote the neighboring nodes of user $u$ and item $i$ respectively, in the unified graph; $\mathbf{e}_{u}^{0}=\mathbf{e}_{u}$ and $\mathbf{e}_{i}^{0}=\mathbf{e}_{i}$.
In particular, the subtraction aggregation operation, as opposed to the addition operation, for aggregating the embeddings of user neighboring nodes to simulate users' inherent intents. 
Finally, the weighted-pooling operation is applied to generate the aggregated representations by operating on the propagated $L$ layers:
\begin{equation}
    \begin{aligned}
    \mathbf{e}_{i}^* = \sum\limits_{l=0}^{L} \alpha_l \mathbf{e}_{i}^l, \quad
    \mathbf{e}_{u}^* = \sum\limits_{l=0}^{L} \alpha_l \mathbf{e}_{u}^l,
    \end{aligned}
\end{equation}
where $\alpha_l$ indicates the importance of the $l$-th layer representation in constituting the final embedding. Following LightGCN~\cite{he2020lightgcn}, we set $\alpha_l$ as $\frac{1}{(l+1)}$, as the focus of our work is not on its selection.

To further constrain the translation relation, we design an intent translation contrastive learning approach that adopts a margin-based ranking criterion. Specifically, we aim to ensure that $\mathbf{e}_u^*+\widetilde{\mathbf{e}}_q\approx \mathbf{e}_i^*$ (i.e., the ground truth interactive item $\mathbf{e}_i^*$ should be near to the translated intent $\mathbf{e}_u^*+\widetilde{\mathbf{e}}_q$), while the negative $\mathbf{e}_{i'}^*$ should be distant from $\mathbf{e}_u^*+\widetilde{\mathbf{e}}_q$, as follows:
\begin{equation}
    \begin{aligned}
        \mathcal{L}_{CL} =\sum\limits_{(u,i,i')\in {Y}}-\ln\sigma[(\mathbf{e}_u^*+\widetilde{\mathbf{e}}_q-\mathbf{e}_{i'}^*)^2- (\mathbf{e}_u^*+\widetilde{\mathbf{e}}_q-\mathbf{e}_i^*)^2],
    \end{aligned}
\end{equation}
where $\widetilde{\mathbf{e}}_q$ denotes the representation of real query in search or the generated demand intent in recommendation for ($u,i$) pair; ${Y}=\{(u,i,i')|(u,i)\in {R}^+,(u,i')\in {R}^-\}$ denotes the pairwise training data where ${R}^+$ indicates the positive observed interaction set, and ${R}^-$ represents the randomly-sampled negative set; $\sigma(\cdot)$ stands for the sigmoid function.  
\subsection{Model Prediction and Optimization}\label{sec:prediction}
After obtaining the representations $\mathbf{e}_{u}^*,\mathbf{e}_{i}^*,\widetilde{\mathbf{e}}_q$, we fuse them to obtain the overall representation for the input sample $x=(u,i,\widetilde{q})$:
\begin{equation}
    \begin{aligned}
        \mathbf{e}_{u,i,\widetilde{q}} =  \mathbf{e}_{u}^* \Vert \mathbf{e}_{i}^* \Vert \widetilde{\mathbf{e}}_q.
    \end{aligned}
\end{equation}

Then, two different MLPs are employed to make prediction for search and recommendation tasks, respectively:
\begin{equation}
    \begin{aligned}
    \hat{y}_{u,i,\widetilde{q}} = \left\{  
             \begin{array}{rcl}
             {\rm MLP_s}(\mathbf{e}_{u,i,\widetilde{q}}) &\mbox{if} & x \in \mathcal{X}_s,\\ 
             {\rm MLP_r}(\mathbf{e}_{u,i,\widetilde{q}}) &\mbox{if} & x \in \mathcal{X}_r.\\
    \end{array} 
\right.  
    \end{aligned}
\end{equation}

We adopt pairwise training to train the model.
Specifically, we adopt the Bayesian Personalized Ranking (BPR)~\cite{rendle2009bpr} loss to emphasize that the observed interaction should be assigned a higher score than the unobserved one as follows:
\begin{equation}
    \begin{aligned}
        \mathcal{L}_{o}=\sum\limits_{(u,i,i')\in {Y}}-\ln \sigma(\hat{y}_{u,i,\widetilde{q}}-\hat{y}_{u,i',\widetilde{q}'}),
    \end{aligned}
\end{equation}
where the representation of $\widetilde{q}'$ denotes the demand intent for the negative pair ($u,i'$). Finally, the overall loss $\mathcal{L}$ is defined using hyper-parameters $\lambda_1$ and $\lambda_2$ as:
\begin{equation}\label{eq:overall_loss}
    \begin{aligned}
    \mathcal{L} = \mathcal{L}_{o}+\lambda_1\mathcal{L}_{SG}+\lambda_2\mathcal{L}_{CL}.
    \end{aligned}
\end{equation}

\section{Experiments}
In this section, we present empirical results to demonstrate the effectiveness of our proposed UDITSR. These experiments are designed to answer the following research questions: \textbf{RQ1} How does UDITSR perform compared with state-of-the-art search and recommendation models?
\textbf{RQ2} What are the effects of the demand intent generator and dual-intent translation mechanism in UDITSR?
\textbf{RQ3} Why could UDITSR perform better?
\textbf{RQ4} How does UDITSR perform in real-world online recommendations with practical metrics?
\textbf{RQ5} How do the hyper-parameters in UDITSR impact the search and recommendation performance?

\subsection{Experimental Settings}

\subsubsection{Dataset Description}
We conducted experiments on two real-world datasets, denoted as MT-Large and MT-Small datasets\footnote{We collected this dataset because there was no public dataset that includes both search and recommendation data. Our code and data will be available at https://github.com/17231087/UDITSR.}. These two datasets are obtained from the Meituan platform, one of the largest takeaway platforms in China. Both datasets span eight days across two cities.
Each sample in the datasets contains a user and an item, and each search sample additionally contains a query.  Specifically, with 111,891 search and 65,035 recommendation interactions collected, our \textbf{MT-Small} dataset comprises 56,887 users and 4,059 items and the average number of split words per query record is 1.6801.  With 1,527,869 search and 1,168,491 recommendation interactions collected, the \textbf{MT-Large} dataset contains 433,573 users and 22,967 items and the average number of split words per query is 1.5561.
To evaluate model performance, we split the first six days’ data for training, the seventh day's data for validation, and the last day's data for testing. 
For each ground truth test record, we randomly sampled 99 items that the user did not interact with as negative samples. 
\begin{table}[H]
  \caption{Network Configuration}\label{tab:network_config}
  \begin{tabular}{c|c}
  \toprule
  Name & Value  \\
  \midrule
  optimizer &AdamW \\
  batch size & 256 \\
  learning rate & 1e-4\\
  weight decay & 1e-5 \\
  vocab size of words in querys &5,000\\
  dimension of embeddings & 100\\
  depth of aggregation & 2\\
  number of words per query   & 3 \\
  number of words per user's historical query & 3 \\
  number of words per item's historical query &  10 \\  
  hidden sizes of ${\rm MLP}$ in demand intent generator & [200,100]\\
  hidden sizes of ${\rm MLP_s}$/${\rm MLP_r}$ & [150,75] \\
  \bottomrule  
  \end{tabular}
\end{table}

\subsubsection{Implementation Details}
We implement all models using PyTorch\footnote{https://pytorch.org/}, a well-known software library for deep learning.  In Section~\ref{sec:hyper_parameter}, we report the impact of essential hyper-parameters in our model, including the loss weights $\lambda_1$ and $\lambda_2$, and we utilize the best settings for these hyper-parameters. The remaining network configurations are presented in Table~\ref{tab:network_config}. 
To ensure a fair comparison, we apply the above-mentioned settings across all models. Moreover, we search for optimal values of the other hyper-parameters of the baseline models as suggested in their respective original papers. Finally, we employ the early stopping strategy based on the models' performance on the validation set to avoid overfitting.

\begin{table*}[]
    \centering
    \begin{tabular}{c|c|cccc||cccc}
    \toprule
    
     \multirow{2}{*}{Dataset}& \multirow{2}{*}{Model} & \multicolumn{4}{c||}{Search}  &  \multicolumn{4}{c}{Recommendation}\\
      \cline{3-10}
       & & Hit@5 &  NDCG@5 & MRR&Avg.C$\downarrow$ &Hit@5 &  NDCG@5 &MRR &AUC  \\
    \midrule
    \multirow{13}{*}{MT-Small}&NeuMF& 0.5510&0.4264&0.4150&12.1907 &0.3147&0.2306&0.2374&0.8160\\
        &DNN& 0.5877&0.4594&0.4465&9.6208&0.3241& 0.2177&0.2246& 0.8150  \\
        &xDeepFM&0.5053&0.3886&0.3815&14.3603  &0.3184&0.2139&0.2218&0.8155\\
        
        &DIN &0.5892  & 0.4726&0.4613&11.6023  &0.3632&0.2510&0.2545& 0.8213  \\
        &AEM &0.5053  &0.3666&0.3568&11.7953&0.3967&0.2703& 0.2686&0.7982 \\
        &TEM &0.5362&0.4185&0.4084&13.6472&0.2933&0.1970&0.2078& 0.7947 \\
        &JSR & 0.6143 &0.4828 &0.4678&8.6276&0.3460&0.2448&0.2457 &0.7532\\
        &SimpleX 
        &0.6237&0.4864&0.4699&8.0841&0.3314&0.2288 &0.2336 &0.8081 \\
        &MGDSPR&0.6150&0.4743&0.4570&8.5362 & 0.2974&0.2032&0.2122&0.7862\\

        \cline{2-10}
        &GAT &0.6025 &0.4679&0.4497 &10.8707&0.4202&0.3109&0.3032&0.7935\\
        &NGCF &0.6418  &0.5173 & 0.5000  &9.7943&0.4564&0.3346&0.3284&\underline{0.8232}\\
        &LightGCN &0.6665&\underline{0.5402}&\underline{0.5195}&9.9139&\underline{0.4577}&0.3296&0.3185&0.8174 \\
        &GraphSRRL&\underline{0.6688}&0.5267&0.5042&\underline{8.0724}& 0.4540&0.3249&0.3159&0.7883 \\
        &SRJgraph &0.6186&0.4850&0.4647 &12.1989 &0.4140&0.3074&0.2997&0.7474 \\
        &DCCF & 0.5013& 0.3760&0.3615&19.6477&0.4323&\underline{0.3380}&\underline{0.3304}&0.7239\\
        \cline{2-10}
        &UDITSR & \textbf{0.7008*}&\textbf{0.5691*}&\textbf{0.5470*}&\textbf{7.5257*}&\textbf{0.4841*}&\textbf{0.3528*}& \textbf{0.3422*}&\textbf{0.8285*}\\
        &Impr.\% &\textbf{4.7847}&\textbf{5.3499}&\textbf{5.2936}&\textbf{6.7725}&\textbf{5.7680}&\textbf{4.3787}&\textbf{3.5714}&\textbf{0.6438}\\
    \midrule
    \midrule
        \multirow{13}{*}{MT-Large}&NeuMF&0.8668& 0.7855 & 0.7682  &3.3001 &0.5390&0.4235&0.4129&0.8573 \\
        &DNN&0.8788&0.7874&0.7664& 3.0520& 0.5153 &0.3962&0.3881&0.8610\\
        
        &xDeepFM&0.8552  &0.7417 &0.7147   &4.0106&0.4926&0.3897&0.3828&0.8077\\
        &DIN &0.8914 &0.7934& 0.7693&2.7283&0.6005&0.4489&0.4292&0.9082\\
        &AEM &0.8760 &0.7654&0.7389&2.9007&0.5865&0.4597& 0.4435&0.8940 \\
        &TEM&0.8611 &0.7522&0.7269 & 3.4096& 0.5031&0.3526&0.3419&0.8899 \\
        &JSR & 0.8691&0.7748&0.7537&3.0903&0.5023&0.3789&0.3683&0.8393\\
        &SimpleX &0.8896 & 0.7895 &0.7651 & 2.6466 &0.5004&0.3790&0.3691&0.8640\\
        &MGDSPR&0.8726& 0.7709&0.7473&3.0325&0.5412&0.4037&0.3888&0.8751 \\
        
        \cline{2-10}
        &GAT &0.8761&0.7796&0.7572&2.9530 &0.5880&0.4540&0.4347&0.8706 \\
        &NGCF &0.8821  &0.7892 &0.7670 &2.7377&\underline{0.6325}&0.4966&0.4780& \underline{0.9096} \\
        &LightGCN &0.8937&\underline{0.8016}&\underline{0.7795}&2.4076 &0.6158& 0.4785&0.4593&0.8920\\
        &GraphSRRL&\underline{0.8966}&0.7992&0.7755&\underline{2.3504}&0.6106&0.4726&0.4543&0.8891 \\
        &SRJgraph &0.8836&0.7883&0.7659 &2.6849 &0.5873&0.4494&0.4315&0.8942 \\
        &DCCF &0.8568&0.7459&0.7205&3.2666&0.6201&\underline{0.5007}&\underline{0.4802}&0.8292\\

        \cline{2-10} 
        &UDITSR &\textbf{0.9178*}&\textbf{0.8382*}&\textbf{0.8183*}&\textbf{1.9819*} &\textbf{0.6566*}&\textbf{0.5157*}&\textbf{0.4936*}&\textbf{0.9146*}\\
        &Impr.\% & \textbf{2.3645}&\textbf{4.5659}&\textbf{4.9775}&\textbf{15.6782}&\textbf{3.8103}&\textbf{2.9958}&\textbf{2.7905}&\textbf{0.5497}\\
    
    \bottomrule
    \end{tabular}
    \caption{Overall performance on both datasets. $\downarrow$ represents that a smaller Avg.C metric value indicates better performance. \textit{Impr.\%} indicates the relative improvements of the best-performing method (bolded) over the strongest baselines (underlined). * indicates 0.05 significance level from a paired t-test comparing UDITSR with the best baselines. }\label{tab:unified_performance}
\end{table*}
\subsubsection{Evaluation Metrics} To evaluate our model's performance, we utilize four widely-used ranking metrics: Hit@K, NDCG@K~\cite{jarvelin2002cumulated} (we set K as 5 by default), MRR~\cite{radev2002evaluating} and Average position of the Clicked items (Avg.C)~\cite{yao2021user}. Additionally, we adopt an accuracy metric, AUC~\cite{ferri2011coherent} for the recommendation task.

\subsubsection{Baselines}
In our work, we evaluate the performance of our model with two groups of baselines to examine its effectiveness.

\noindent\textit{(1) Graph-free baselines}

\begin{itemize}[leftmargin=*,itemsep=2pt,topsep=0pt,parsep=0pt]
    \item \textbf{NeuMF}~\cite{he2017neural} combines traditional matrix decomposition with the MLP to extract low-dimensional and high-dimensional features simultaneously.
    \item \textbf{DNN} combines the embedding layer described in Section~\ref{sec:embedding_layer} with the prediction layer described in Section \ref{sec:prediction}. 
    \item \textbf{xDeepFM}~\cite{lian2018xdeepfm}  consists of a compressed interaction network (CIN)  and an MLP for prediction, where CIN generates explicit feature interactions at the vector-wise level.
    \item \textbf{DIN}~\cite{zhou2018deep} utilizes an attention mechanism between the historical behavior sequence and the target item to model the evolving interests. 
    \item \textbf{AEM}~\cite{ai2019zero} allocates different attention values to the previous behavior sequence based on the current search queries. 
    \item \textbf{TEM}~\cite{bi2020transformer} feeds the sequence of query and user behavior history into a transformer layer to extract the search intents.
    \item \textbf{JSR}~\cite{zamani2020learning} integrates neural collaborative filtering and language modeling to reconstruct query text descriptions, enabling the joint model of search and recommendation.
    \item \textbf{SimpleX}~\cite{mao2021simplex}  is a simplified variant of the two-tower model with user behavior modeling. 
    \item \textbf{MGDSPR}~\cite{li2021embedding} utilizes an attention mechanism to model the relationship between users' query multi-grained semantics and their personalized behaviors for prediction.

\end{itemize}

\noindent\textit{(2) Graph-based baselines}
\begin{itemize}[leftmargin=*,itemsep=2pt,topsep=0pt,parsep=0pt]
    \item \textbf{GAT}~\cite{velivckovic2018graph} utilizes the attention mechanism to measure the importance of neighbor nodes during the aggregation process.
    \item \textbf{NGCF}~\cite{wang2019neural} enhances the Graph Convolutional Networks (GCN) by incorporating user-item interactions.
    \item \textbf{LightGCN}~\cite{he2020lightgcn} streamlines GCN by relying solely on neighborhood aggregation to capture collaborative filtering, omitting feature transformation and non-linear activation components.
    \item \textbf{GraphSRRL}~\cite{liu2020structural} exploits three specific structural patterns within a user-query-item graph.
    \item \textbf{SRJGraph}~\cite{zhao2022joint} incorporates padding queries for recommendation and search queries as attributes into interaction edges, enabling joint modeling of both tasks.
    \item \textbf{DCCF}~\cite{ren2023disentangled} leverages an adaptive self-supervised augmentation to disentangle intents behind user-item interactions. 
   
\end{itemize}
Specifically, NeuMF, xDeepFM, DIN, DCCF, SimpleX, NGCF and LightGCN are proposed for the \textbf{recommendation task}, while AEM, TEM, MGDSPR and GraphSRRL are proposed for the \textbf{search task}. JSR and SRJGraph are designed for \textbf{joint learning of both tasks}. To adapt these baselines for both tasks, \textbf{real query representations for search and padding query representations for recommendation are incorporated into the prediction layer} described in Section \ref{sec:prediction}. Previous studies~\cite{zamani2018joint,zhao2022joint} have demonstrated that joint optimization of search and recommendation models can improve performance, so all baselines are directly trained on both search and recommendation data. 
All baselines use the same settings for the embedding layer and the prediction layer, and the interaction graph is built on both search and recommendation interactions.

\subsection{Overall Performance Comparison (RQ1)}\label{sec:overall_performance}

We present the results on the two adopted datasets in Table~\ref{tab:unified_performance}.  From the results, we can observe that:
\begin{itemize}[leftmargin=*,itemsep=2pt,topsep=0pt,parsep=0pt]
    \item  UDITSR significantly outperforms all the competitive baselines on both tasks. Specifically, compared to the best-performing baselines, UDITSR gains an average improvement of 6.22\% and 3.06\%  in the search and recommendation tasks, respectively. 
    \item Most graph-based methods, such as NGCF, LightGCN, and GraphSRRL, perform well in both tasks, potentially due to their ability to effectively capture complex high-order interactive patterns.  
    \item SRJgraph assumes that query-related intents in recommendation remain unchanging whereas in search, the matching degree between the query and the candidate items is deemed crucial. Consequently, such an assumption may limit the model's performance, particularly when compared to our UDITSR, which learns and adapts to changing query-related intents. 
    
\end{itemize}

\subsection{Ablation Study (RQ2)}\label{sec:core_component}
As the demand intent generator and dual-intent translation propagation are the core of our model, we conduct the following ablation studies to investigate their effectiveness: 
\begin{itemize}[leftmargin=*,itemsep=2pt,topsep=0pt,parsep=0pt]

    \item UDITSR(w/o DeIntGen) masks all generated demand intents $\widetilde{\mathbf{e}}_q$ by assigning the embedding of the padding query to each recommended record.
    \item UDITSR(w/o IntTrans) replaces dual-intent translation with classical mean-pooling propagation between the user and item nodes. 
    \item UDITSR(w/o DeIntGen \& IntTrans) removes both the demand intent generator and dual-intent translation propagation, as described in the two ablation studies above.
\end{itemize}
\begin{table*}[]\small
    \centering
    \begin{tabular}{c|c|cccc||cccc}
    \toprule
      \multirow{2}{*}{Dataset}&\multirow{2}{*}{Ablation} & \multicolumn{4}{c||}{Search}  &  \multicolumn{4}{c}{Recommendation}\\
      \cline{3-6}\cline{7-10}
      & & Hit@5 &  NDCG@5 & MRR&Avg.C$\downarrow$& Hit@5 &  NDCG@5 &MRR &AUC \\
    
    \midrule
        \multirow{4}{*}{MT-Small} &UDITSR(w/o DeIntGen \& IntTrans) & 0.6479  &0.5152 & 0.4949 &10.1960  &0.4185 &0.3052 &  0.3032&0.8179 \\
        &UDITSR(w/o IntTrans) & 0.6454&0.5130&0.4924 &10.5527&0.4352&0.3206& 0.3154&0.8225 \\
        &UDITSR(w/o DeIntGen) &0.6959&0.5543&0.5307&8.1389&0.4510&0.3237&0.3178&0.8186
        \\
        
       &UDITSR &\textbf{0.7008*}&\textbf{0.5691*}&\textbf{0.5470*}&\textbf{7.5257*}&\textbf{0.4841*}&\textbf{0.3528*}& \textbf{0.3422*}&\textbf{0.8285*}\\
        \midrule
        
        \midrule
         \multirow{4}{*}{MT-Large}&UDITSR(w/o DeIntGen \& IntTrans) &0.8660 &0.7586 &0.7337 &3.1185 &  0.6183&0.4818&0.4623  &0.9031  \\
        &UDITSR(w/o IntTrans) &0.8870&0.7866&0.7623&2.6399 &0.6228&0.4879&0.4696&0.9061\\
        &UDITSR(w/o DeIntGen) &0.9089 &0.8192&0.7969&2.2053 &0.6303&0.4890& 0.4685&0.9058\\
        &UDITSR &\textbf{0.9178*}&\textbf{0.8382*}&\textbf{0.8183*}&\textbf{1.9819*} &\textbf{0.6566*}&\textbf{0.5157*}&\textbf{0.4936*}&\textbf{0.9146*}\\
        
        \bottomrule
    \end{tabular}
    \caption{Ablation study on our proposed search-supervised demand intent generator and dual-intent translation propagation.}
    \label{tab:ablaton study}
\end{table*}

From the results of ablation studies in Table ~\ref{tab:ablaton study}, we can find that: 
\begin{itemize}[leftmargin=*,itemsep=2pt,topsep=0pt,parsep=0pt]
    \item UDITSR(w/o DeIntGen \& IntTrans) performs the worst on both search and recommendation tasks, suggesting that the significant improvement of our model stems from our proposed demand intent generator and dual-intent translation propagation.  
    \item UDITSR(w/o IntTrans) performs worse than the original UDITSR, highlighting the effectiveness of our proposed intent translation propagation mechanism. 
    \item  UDITSR(w/o DeIntGen) performs worse than UDITSR, especially for recommendation task, indicating that the search-supervised demand intent generator can help UDITSR learn implicit intents more accurately in recommendation.
\end{itemize}

\begin{figure}
    \centering 
    \subfigure[Intents in UDITSR(w/o IntTrans) for search data]
    {
    \includegraphics[width=0.3\linewidth]{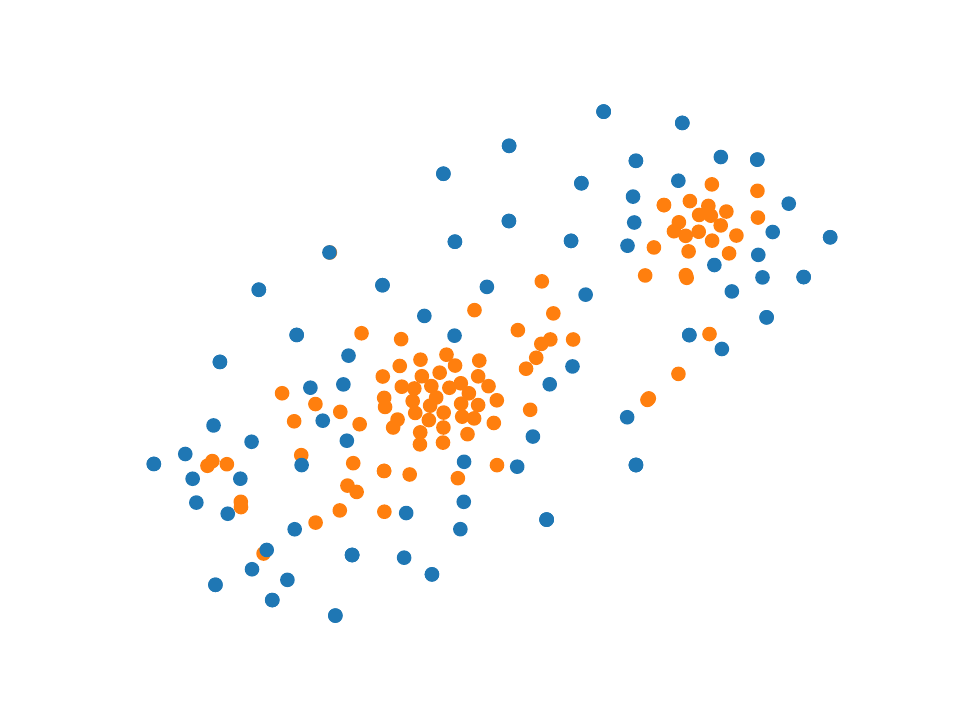}
    }
    \subfigure[Inherent intents in UDITSR for search data]
    {
    \includegraphics[width=0.3\linewidth]{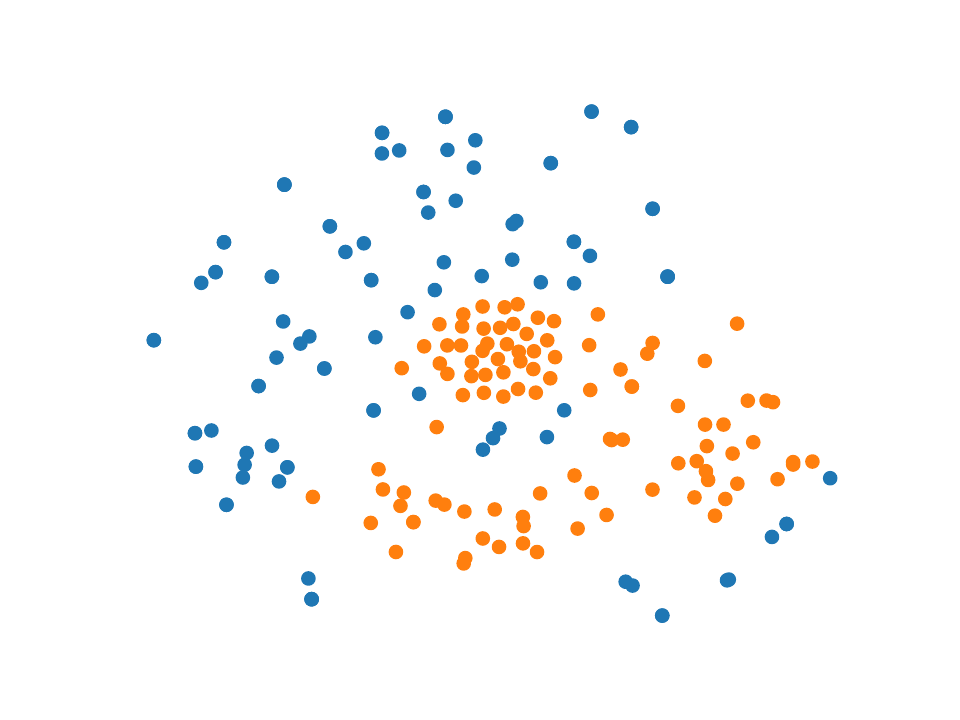}
    }
    \subfigure[Translated intents in UDITSR for search data]
    {
    \includegraphics[width=0.3\linewidth]{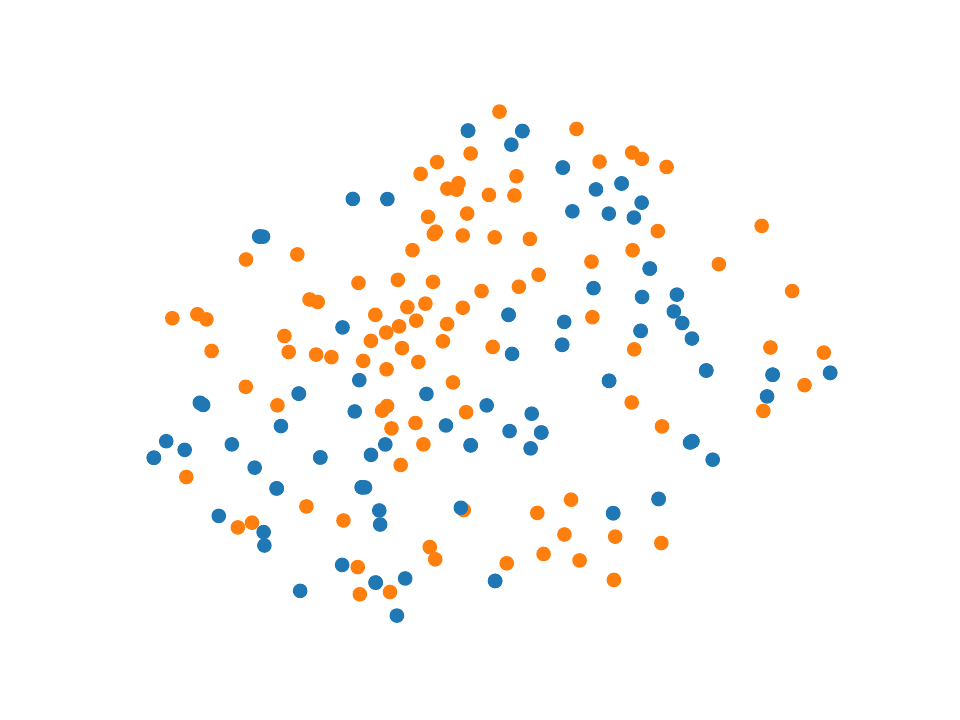}
    }
    \subfigure[Intents in UDITSR(w/o IntTrans) for recommendation data]
    {
    \includegraphics[width=0.3\linewidth]{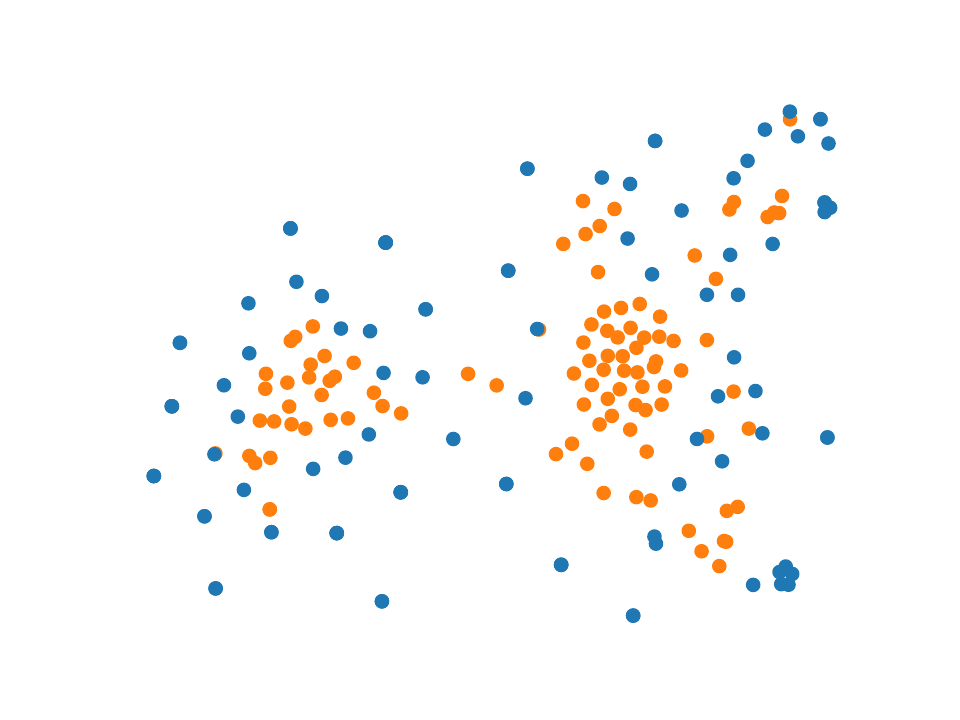}
    }
    \subfigure[Inherent intents in UDITSR for recommendation data]
    {
    \includegraphics[width=0.3\linewidth]{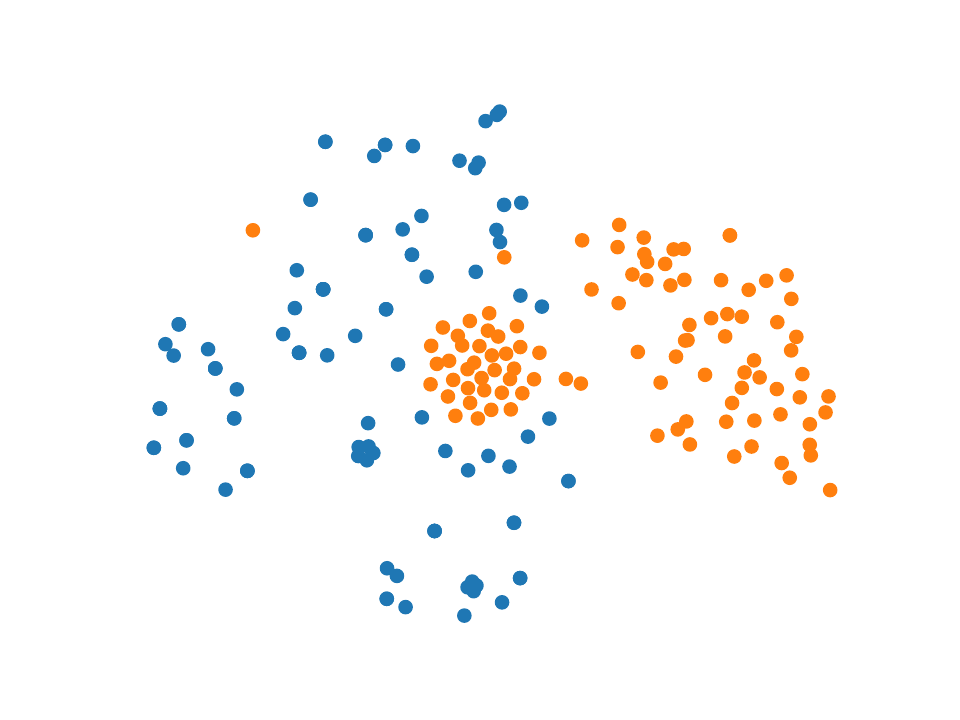}
    }
    \subfigure[Translated intents in UDITSR for recommendation  data]
    {
    \includegraphics[width=0.3\linewidth]{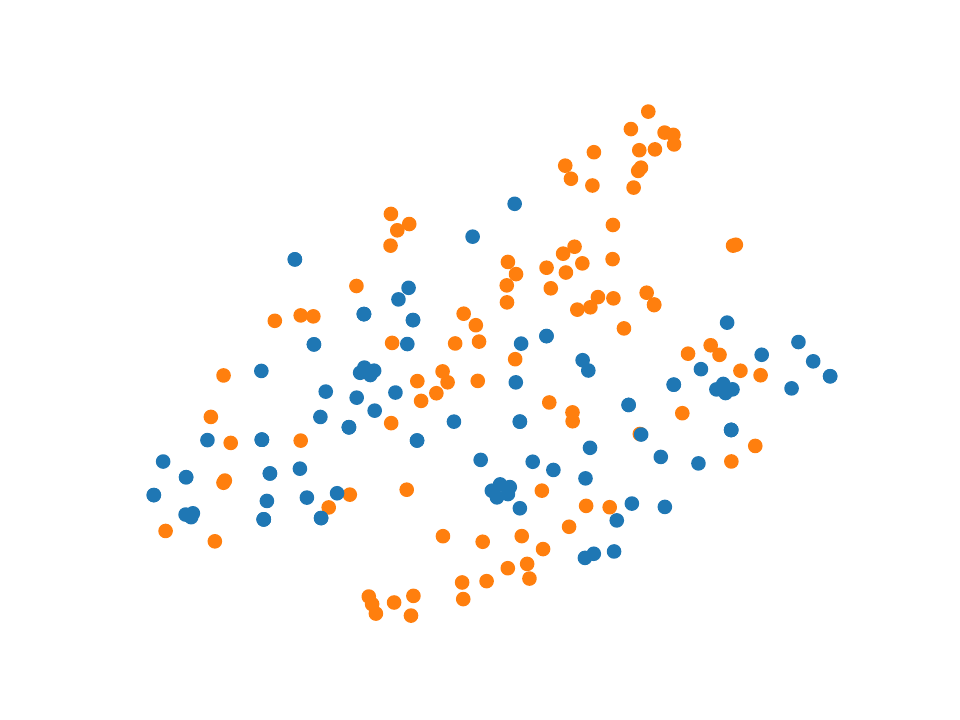}
    }
    \caption{t-SNE visualization of learned intents and interactive items. Blue dots represent the interactive items(i.e., $\mathbf{e}_i^*$) and orange dots represent the learned intents.}\label{fig:intent_translation_dis}

\end{figure}

\subsection{Intent Visualization (RQ3)}
In this section, we visualize the learned intents to further investigate why our model performs better. We compare UDITSR with its ablated version without the dual-intent translation propagation (UDITSR(w/o IntTrans), detailed in Section~\ref{sec:core_component}).  We employ the default setting of the t-SNE~\cite{donahue2014decaf} provided by Scikit-learn to visualize the distribution of the learned intents and the interactive items. 
For clarity, we randomly sample 100 positive records from the search and recommendation test datasets respectively for plotting. Specifically, in UDITSR(w/o IntTrans), user embeddings (i.e.,$\mathbf{e}_u^*$) are regarded as the learned intents, as shown in Figures~\ref{fig:intent_translation_dis}(a) and (d), similar to preference/intent captured by models like NGCF and LightGCN. UDITSR, however, couples inherent and demand intents via intent translation to form the final intents (i.e., $\mathbf{e}_u^*+\widetilde{\mathbf{e}}_q$), as shown in Figure~\ref{fig:intent_translation_dis}(c) and (f). To ensure a fair comparison, we present the inherent intents (i.e., $\mathbf{e}_u^*$) learned by UDITSR  in Figure ~\ref{fig:intent_translation_dis}(b) and (e). 

Ideally, the distribution of learned intents should match  that of interactive item representations. Figures~\ref{fig:intent_translation_dis}(a) and (d) reveal that the intents learned by UDITSR(w/o IntTrans) are concentrated while the positive interactive items are scattered, indicating a mismatch. Meanwhile, the inherent intents learned by UDITSR are relatively scattered, indicating that our model can better learn the personalized inherent intents of different users. However, there still exist obvious gaps between the intents and items, highlighting the necessity of learning demand intents. In contrast, the translated intents learned by UDITSR are scattered in the space of the target interactive items, demonstrating its excellent intent modeling capability. The better fit of the distribution of translated intents to the target interactive distribution could be the fundamental reason for the better overall performance of UDITSR.

\begin{figure}
    \centering 
    \includegraphics[width=\linewidth]{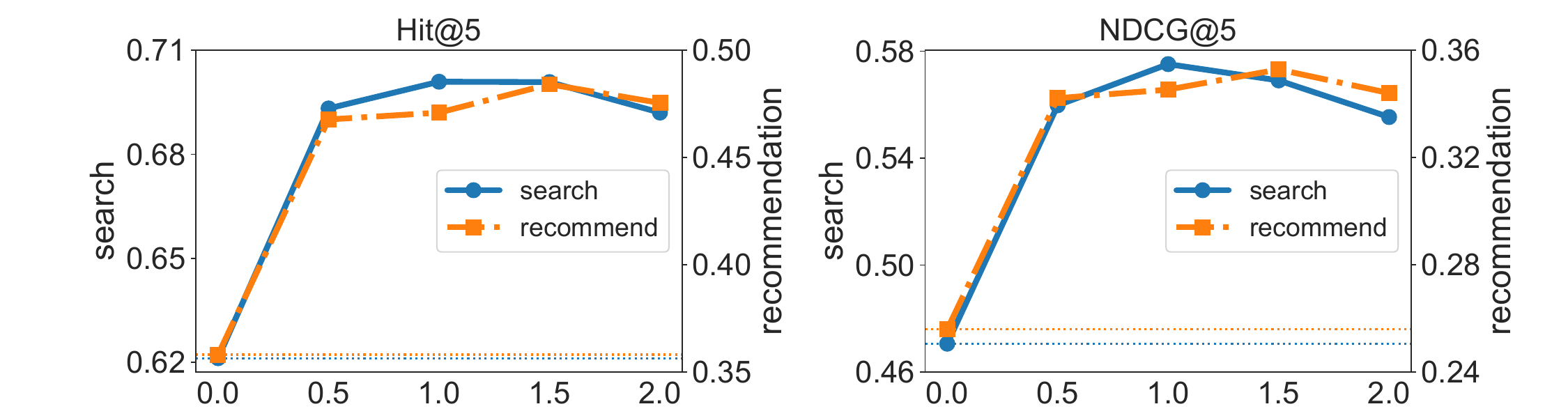}
    \caption{Performance w.r.t $\lambda_1$ of search-supervised demand intent generator for search and recommendation tasks.}\label{fig:search_lambda}
\end{figure}
\begin{figure}
    \centering 
    \includegraphics[width=\linewidth]{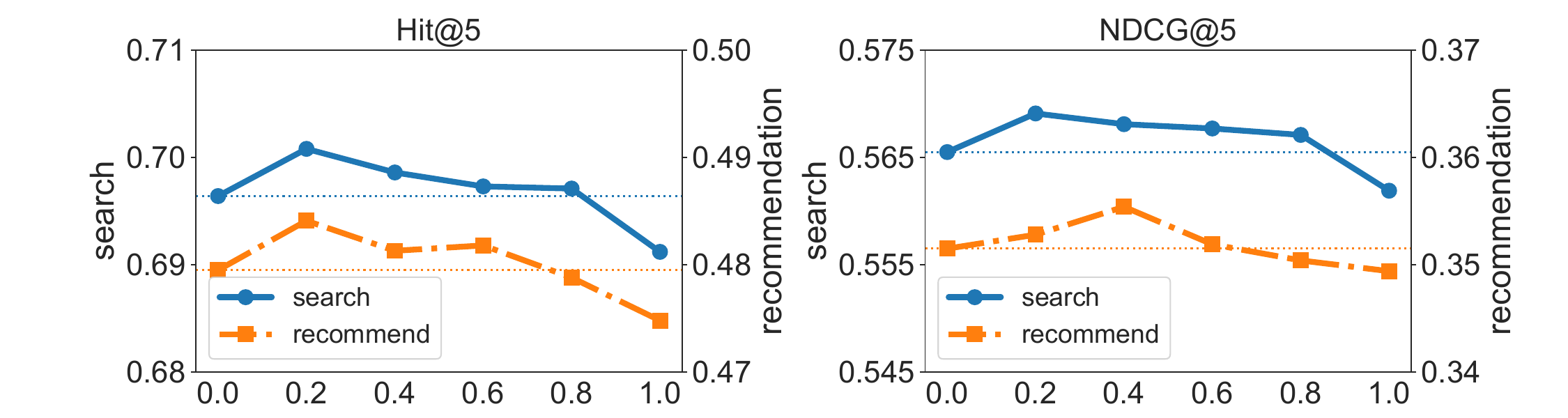}
    \caption{Performance w.r.t $\lambda_2$ of the intent translation contrastive learning for search and recommendation tasks.}\label{fig:cst_lambda}
\end{figure}

\subsection{Online A/B test (RQ4)}
Owing to the distinct architectural differences between the search and recommender systems on the Meituan Waimai platform, we have initially focused our methodological deployment on the homepage recommender systems. We conducted a month-long online A/B test from December 18, 2023, to January 17, 2024. Specifically, we utilized the search data with query information to guide the learning of user demand intent representation and leveraged the learned graph embeddings as additional features in the downstream recommendation model. The control bucket was the original online recommendation method of Meituan Waimai platform. The deployment of our method increased the \textbf{GMV}(Gross Merchandise Volume) by \textbf{1.46}\% and \textbf{CTR}(Click-Through Rate) by \textbf{0.77\%}, which demonstrated the effectiveness of our method. In the future, we will continue to conduct comprehensive online experiments that encompass both search and recommendation scenarios.

\subsection{Hyper-Parameter Studies (RQ5)}\label{sec:hyper_parameter}
In this section, we conduct experiments on the loss weights ($\lambda_1$, $\lambda_2$) in Eq.~\ref{eq:overall_loss} on MT-Small dataset to explore their impact. 

(1) Loss weight of the demand intent generator ($\lambda_1$). 
We vary $\lambda_1$ within $\{0,0.5,1.0,1.5,2.0\}$. The results in Figure~\ref{fig:search_lambda} indicate that performance improves and then declines with increasing $\lambda_1$. With $\lambda_1 =0$, the demand intent generator degenerates to an ordinary generator without any search-supervision information. All models with search supervision (i.e. $\lambda_1 \neq 0$) outperform models without it (i.e. $\lambda_1= 0$). 
This may stem from \textbf{UDITSR's effective learning of user demand intents through explicit supervision from search}. 
Furthermore, our model excels across most metrics for both search and recommendation tasks at $\lambda_1=1.5$. Thus, we set $\lambda_1=1.5$ for MT-Small dataset. After a similar experiment conducted on MT-Large dataset, we adopt the best-performing setting ($\lambda_1=1$).

(2) Loss weight of the intent translation contrastive learning ($\lambda_2$).
To investigate the impact of our proposed intent translation contrastive learning, we vary $\lambda_2$ in $\{0.0,0.2,0.4,0.6,0.8,1.0\}$. 
Overall,  the performance initially increases and then decreases with the increase of $\lambda_2$. Particularly,  our model with $\lambda_2$ set in \{0.2, 0.4, 0.6\} outperforms the version without translation contrastive learning $\lambda_2=0$ on all metrics, \textbf{demonstrating a proper loss weight of intent translation contrastive learning can aid in intent relation modeling}. The optimal $\lambda_2$ for search is 0.2 while for recommendation task, it is  $0.2$ for Hit@5 and $0.4$ for NDCG@5. Therefore, we set $\lambda_2=0.2$ for MT-Small dataset. Also, after conducting a similar experiment on MT-Large dataset,  we adopt the best-performing setting $\lambda_2=0.4$.

\section{Conclusion}
This paper introduced a novel approach to unified intention-aware modeling for joint optimization of search and recommendation tasks. We recognized that user behaviors were motivated by their inherent intents and changing demand intents. To accurately learn users' implicit demand intents for recommendation, we innovated a demand intent generator that utilized explicit queries from search data for supervised learning. Furthermore, we proposed a dual-intent translation propagation mechanism for interpretive modeling of the relation between users' dual intents and their interactive items. 
In particular, we introduced an intent translation contrastive method to further constrain this relation. Our extensive offline experiments demonstrated that UDITSR outperformed the leading baselines in both search and recommendation tasks. Besides, online A/B tests further confirmed the superiority of our model.  
Finally, the intent visualization clearly explained the deeper reason for the remarkable improvement of our model.

\begin{acks}
This research work is supported by the National Key Research and Development Program of China under Grant No.2021ZD0113602, the National Natural Science Foundation of China under Grant No.62176014 and No.62306255, the Fundamental Research Funds for the Central Universities and the Fundamental Research Project of Guangzhou under Grant No. 2024A04J4233.
\end{acks}

\bibliographystyle{ACM-Reference-Format}
\balance
\bibliography{ref}

\end{document}